\documentstyle[aps,prl,epsf,epsfig,floats,twocolumn]{revtex}
\begin{document}
\draft
\twocolumn[\hsize\textwidth\columnwidth\hsize\csname @twocolumnfalse\endcsname

\title{Magnetooptical sum rules close to the Mott transition}
\author{Ekkehard Lange and Gabriel Kotliar}
\address{Serin Physics Laboratory,
Rutgers University,
136 Frelinghuysen Road,
Piscataway, New Jersey 08854, USA
}
%\date{\today}
\maketitle
\begin{abstract}

We derive new sum rules for the real and imaginary parts of the
frequency-dependent Hall constant and Hall conductivity. As an
example, we discuss their relevance to the doped Mott insulator that
we describe within the dynamical mean-field theory of strongly
correlated electron systems.

\end{abstract}

%%
%%78.20.Ls	Magnetooptical effects
%%72.15.Gd	magnetotransport effects
%%74.25.Gz	Optical properties
%%71.27.+a	Strongly correlated electron systems
%%
]

The ac Hall effect can provide  valuable insights  into the dynamics of
an electronic medium. This has recently been demonstrated in the case
of high-$T_c$ superconductors \cite{Spielman:1994,Kaplan:1996}:
Various theoretical models based on different scattering mechanisms
agree that the anomalous frequency and temperature dependences of the
Hall effect are closely intertwined, but they differ in their
predictions about these dependences \cite{Zheleznyak:1998}. So far,
experiments cannot discriminate between these models, but they will
possibly be able to do so in the future \cite{Zheleznyak:1998}. 

The magnetooptical response of charge carriers can be probed by
the frequency-dependent Hall conductivity, Hall constant, or Hall
angle. Recently, a sum rule for the Hall angle has been derived
\cite{Drew:1997} that is similar to the well-known $f$-sum rule for
the optical conductivity \cite{Nozieres:1958}. In this paper, we
derive new sum rules for the real and imaginary parts of the two other
magnetotransport probes. Such sum rules are useful: First, they help
elucidating how the corresponding spectral weight is redistributed
upon changing the temperature or the doping level. Second, they
provide exact constraints on the interdependence of Hall
effect-related quantities and thus help interpreting experimental
data. For example, the sum rules for the ac Hall constant relate its
low-frequency behavior to its infinite-frequency limit. This can be
useful because experimentally, only the microwave domain and the far
infrared are attainable sufficiently reliably
\cite{Kaplan:1996,Zheleznyak:1998}, whereas the calculation of the
Hall constant simplifies considerably in the high-frequency limit
\cite{Shastry:1993}.
 
We shall first derive the sum rules for the Hall conductivity and Hall
constant quite generally. Then, to illustrate their application, we
shall discuss some aspects of the magnetooptical response of
correlated electrons close to the density-driven Mott transition.

We start by considering the ac conductivities. In terms of the
dissipative part of the current-current correlation function,
\begin{equation}
 \chi_{\nu\mu}''(\omega)=\int_{-\infty}^{\infty}  dt\,
   {1\over2}\langle[\hat{J}_{\nu}(t),\hat{J}_{\mu}(0)]\rangle e^{i\omega t},
\label{chi''}
\end{equation}
the conductivity tensor reads
\begin{equation}
 \sigma_{\nu\mu}(\omega)=ie^2{\cal P}\int_{-\infty}^{\infty}
    {d\tilde{\omega}\over\pi}\frac{\chi_{\nu\mu}''(\tilde{\omega})}
    {\tilde{\omega}(\omega-\tilde{\omega})}
    +e^2\frac{\chi_{\nu\mu}''(\omega)}{\omega}\label{condten}.
\end{equation}
Here, ${\cal P}$ indicates principal-value integration. From time
reversal invariance, homogeneity of time, and the Hermiticity of the 
current operators, we may deduce the following symmetry properties
\cite{Lange:1997a}:
\begin{eqnarray}
 \chi_{xx}''(\omega)&=&\mbox{odd \& real}\label{chixx}\\
 \chi_{xy}''(\omega)&=&\mbox{even \& wholly imaginary}\label{chixy},
\end{eqnarray}
where Eq.\ (\ref{chixy}) holds to first order in the magnetic field. 
Eqs.\ (\ref{chixx}) and (\ref{chixy}) imply that the real parts of 
$\sigma_{xx}(\omega)$ and $\sigma_{xy}(\omega)$ are even while their
imaginary ones are odd. We also see that the dc Hall conductivity is
finite only if 
\begin{equation}
 \chi_{xy}''(0)=0,
\label{s_H(0)}
\end{equation}
and derive the first couple of sum rules:
\begin{eqnarray}
 \int_0^{\infty}d\omega\,\Re\sigma_{xy}(\omega)&=&0,\label{suhconre}\\
 \int_{-\infty}^{\infty}d\omega\,\frac{\omega\Im\sigma_{xy}(\omega)}
   {\pi e^2}&=&-i\langle[\hat{J}_x,\hat{J}_y]\rangle\label{suhconim}.
\end{eqnarray}
To prove Eq.\ (\ref{suhconre}), we close the path of integration along
a semicircle at infinity in the upper-half complex-frequency ($z$)
plane and apply Cauchy's theorem. The integral on the semicircle does
not contribute since the leading high-frequency behavior of
$\sigma_{xy}(z)$ is $1/z^2$. The sum rule (\ref{suhconim})
is similar to the $f$-sum rule of the optical conductivity,
\begin{equation}
 \int_{-\infty}^{\infty}d\omega\,\frac{\Re\sigma_{xx}(\omega)}{\pi
	e^2}=i\langle[\hat{J}_x,\hat{P}_x]\rangle=\chi^0,
\label{f-sumr}
\end{equation}
where $\hat{P}_x$ is the polarization operator satisfying
$\hat{J}_x(t)=\partial\hat{P}_x(t)/\partial t$, and  
$\chi^0=\int d\omega\,\chi_{xx}''(\omega)/\pi\omega$ is the
static current-current correlation function, which is positive
definite. To interprete the right-hand sides of Eqs.\ (\ref{suhconim})
and (\ref{f-sumr}), we first note that the Hall frequency
$\omega_H\equiv-i\langle[\hat{J}_x,\hat{J}_y]\rangle/\chi^0$ is the
generalization of the cyclotron frequency to the lattice
\cite{Lange:1997a,Drew:1997}. Its sign determines that of the
infinite-frequency Hall constant,
\begin{equation}
 R_H^*=\lim_{H\rightarrow0}\frac{N\omega_H}{e^2\chi^0H},
\label{Ihall}
\end{equation}
which was considered by Shastry {\it et al.} \cite{Shastry:1993}. 
Here, $N$ denotes the total number of lattice sites.
Second, the Drude-theory expression
$\sigma_{xx}(\omega)=\frac{\omega_p^2/4\pi}{1/\tau-i\omega}$ yields
$\chi^0=\omega_p^2/4\pi e^2$, where $e$ is the charge of an electron
and $\omega_p$ the plasma frequency. In general, however, $\chi^0$ and
$\omega_H$ depend on all external and model parameters such as
temperature, band filling, and correlation strength.

Before proceeding, we compare the sum rules (\ref{suhconim}) and
(\ref{f-sumr}). In both cases, the contribution of a band
$\epsilon(\vec{k})$ to the right-hand side can be represented as a
weighted average of the momentum-distribution function,
$n_{\vec{k}\sigma}$, over the Brillouin zone (BZ), where the weight
function is determined by the inverse mass tensor
\cite{Shastry:1993}:
$-i\langle[\hat{J}_x,\hat{J}_y]\rangle=He\sum_{\vec{k}\sigma}
\det(\epsilon^{\nu\mu}_{\vec{k}})n_{\vec{k}\sigma}$ and
$\chi^0=\sum_{\vec{k}\sigma}\epsilon^{xx}_{\vec{k}}n_{\vec{k}\sigma}$.
Here, upper indices indicate differentiation with respect to a
component of the Bloch vector, such as in, say,
$\epsilon_{\vec{k}}^x={\partial\epsilon_{\vec{k}}/\partial k_x}$. $H$
is the magnetic field and is assumed to point in the $z$ direction,
and $\nu,\mu=x,y$. In many semiconductors, only Bloch states close to
the minima of the conduction band or the maxima of the valence band
contribute. Then, one can replace the inverse mass tensor by its value at
the respective band edge. Thus, the sum rules (\ref{suhconim}) and
(\ref{f-sumr}) are seen to relate hard-to-obtain experimental
information to, first, the number of carriers and, second, to the mass
tensor at a band edge which can be measured in a cyclotron-resonance
experiment. In a strongly correlated system, on the other hand, the
momentum-distribution function receives contributions from the entire 
BZ, and the above-mentioned BZ averages may no longer be easy to
determine experimentally \cite{Shastry:1993}.

Next, we investigate the ac Hall constant. In Ref.\
\cite{Lange:1997a}, it has been decomposed into its infinite-frequency
limit (\ref{Ihall}) and a memory-function contribution which can be
represented in terms of a spectral function $k(\omega)$:
\begin{equation}
 R_H(\omega)=R_H^*\left(1+\int_{-\infty}^{\infty}
 	d\tilde{\omega}\,{\cal P}
	\frac{k(\tilde{\omega})\tilde{\omega}}
	{\tilde{\omega}-\omega}\right)
	+i\pi R_H^*k(\omega)\omega.
\label{acRH}
\end{equation}
$k(\omega)$ was shown to be even and real. Therefore, the real and
imaginary parts of $R_H(\omega)$ are even and odd, respectively. We
also establish nontrivial sum rules for the ac Hall constant:
\begin{eqnarray}
 \int_0^{\infty}d\omega\,\left[\Re R_H(\omega)-R_H^*\right]=0,
\label{acHallre}\\
 \int_{-\infty}^{\infty}\frac{d\omega}{\pi}\,\frac{\Im
 R_H(\omega)}{\omega}=R_H-R_H^*,
\label{acHallim}
\end{eqnarray}
where $R_H$ is the dc Hall constant. Eq.\ (\ref{acHallre}) holds
because the leading high-frequency behavior of $R_H(z)-R_H^*$ is
$1/z^2$ \cite{Lange:1997a}. Eq.\ (\ref{acHallim}) is a Kramers-Kronig
relation. The sum rules (\ref{acHallre}) and (\ref{acHallim}) are
interesting because they relate the Hall constant at low frequencies
to its infinite-frequency limit. The low-frequency regime is
attainable in experiments \cite{Kaplan:1996}, whereas the
high-frequency limit is much easier to handle theoretically. The sum
rule (\ref{acHallre}) implies that $R_H$ cannot go over from its dc
value to its infinite-frequency limit monotonically. 

Finally, we quote a sum rule for the Hall angle $t_H(\omega)\equiv
\tan\theta_H(\omega)=\sigma_{xy}(\omega)/\sigma_{xx}(\omega)$
that was derived in Ref.\ \cite{Drew:1997}:
\begin{equation}
 \int_{-\infty}^{\infty}\frac{d\omega}{\pi}\,\Re t_H(\omega)=\omega_H.
\label{suhang}
\end{equation}

By contrast to the $f$-sum rule in Eq.\ (8), none of our sum rules 
involves a positive definite integrand. As a consequence, we expect
our sum rules to become fully useful only in conjunction with some
theoretical understanding of the problem involved.

We now apply the above-mentioned sum rules to the doped Mott
insulator, which we describe by the single-band Hubbard model with
bare bandwidth $2D$ and on-site repulsion $U$. We are primarily
interested in the physics close to half filling,
$\delta\equiv1-n\ll1$, where $n$ denotes the average occupancy per
lattice site. In the limit of infinite spatial dimensions, all vertex
corrections of the conductivity tensor vanish which implies
\cite{Voruganti:1992} 
\begin{eqnarray}
 \sigma_{xx}(i\omega_m)&=&{2ie^2\over N\beta}\sum_{\vec{k}n}
	(\epsilon_{\vec{k}}^x)^2
 	G_{\vec{k}|n}\,\frac{G_{\vec{k}|n+m}-G_{\vec{k}|n}}{i\omega_m},
\label{ocond}\\ 
 \sigma_{xy}(i\omega_m)&=&{e^3H\over N\beta}\sum_{\vec{k}n}
   \left|\begin{array}{cc}
	\epsilon_{\vec{k}}^x\epsilon_{\vec{k}}^x &
	\epsilon_{\vec{k}}^{xy}\\[1.4 ex]
	\epsilon_{\vec{k}}^y\epsilon_{\vec{k}}^x &
	\epsilon_{\vec{k}}^{yy}
   \end{array}\right|
   G_{\vec{k}|n}^2\,\frac{G_{\vec{k}|n+m}-G_{\vec{k}|n-m}}{i\omega_m}.
\label{hcond}
\end{eqnarray}
In each equation, a spin factor 2 has been taken into account and  
$\beta=1/T$ is the inverse temperature. $i\omega_n$ and $i\omega_m$
are fermionic and bosonic Matsubara frequencies, respectively. The
single-particle Green's function is given by
$G_{\vec{k}|n}^{-1}=i\omega_n+\mu-\epsilon_{\vec{k}}-\Sigma(i\omega_n)$,
where the local self-energy $\Sigma(i\omega_n)$ must be calculated by
solving a single-impurity Anderson model supplemented by a
self-consistency condition \cite{Georges:1996}. Earlier work on the
Hall effect in infinite dimensions was carried out in Refs.\
\cite{Pruschke:1995,Majumdar:1995,Lange:1998a}. We compute the ac
conductivities (\ref{ocond}) and (\ref{hcond}) numerically by using
the tight-binding band $\epsilon_{\vec{k}}=-(D/\sqrt{2d})\sum_j
\cos(k_ja)$ in $d$ dimensions, where $a$ is the lattice spacing. We 
use the iterated perturbation theory (IPT), which can be shown to obey
our sum rules exactly.

Our main focus is on the frequency regime well below the Mott-Hubbard
gap $U$. The relevant part of the single-particle spectrum then
consists of two distinct features: an incoherent lower Hubbard band
(LHB) and, provided the temperature is low enough, a quasiparticle
resonance (QPR) at the Fermi level. As the doping level is increased,
the QPR merges with the LHB from above.

Accordingly, there are two widely different energy scales close to the
Mott transition: a coherence temperature $T_{\mbox{\small coh}}$
below which Fermi-liquid properties begin to be observed, and $D$
which sets the scale for incoherent excitations. The width of the QPR
defines a second low-energy scale $T^*$. The ac conductivities
(\ref{ocond}) and (\ref{hcond}) reflect the possible transitions
within the single-particle spectrum. For $T<T_{\mbox{\small coh}}$,
this means that the integrands of all sum rules roughly decompose into
two features: First, a narrow one at zero frequency which is due to
transitions within the QPR. Consequently, its width scales at most
with $T^*$. We shall see below that this feature can be resolved in
the Fermi-liquid regime so its width does not exceed the smaller scale
$T_{\mbox{\small coh}}$. Second, a feature around a frequency
$\omega_1$ that measures the distance between the maxima of the LHB
and the QPR, $\omega_1\sim D$. At high temperatures, on the other
hand, the integrands are solely determined by transitions from
occupied to unoccupied states within the LHB, and $D$ is the only
energy scale.

In the Fermi-liquid regime, $T,\omega<T_{\mbox{\small coh}}$,
the conductivities can be cast into the Drude forms 
$\sigma_{xx}(\omega)=\frac{\omega_p^{*2}/4\pi}{1/\tau^*-i\omega}$
and $\sigma_{xy}(\omega)=\frac{\omega_c^*\omega_p^{*2}/4\pi}
{(1/\tau^*-i\omega)^2}$ with renormalized parameters. Here, the
renormalized plasma frequency behaves as $\omega_p^{*2}\sim D\delta$
\cite{Rozenberg:1996};
$1/\tau^*\sim\delta\,\mbox{Im}\Sigma_R(\omega=0,T)$, 
where $\Sigma_R$ is the retarded self-energy in the absence of
disorder; $\omega_c^*\sim\omega_c\delta$ where $\omega_c$ is the
cyclotron frequency of noninteracting electrons on the  same
lattice. The renormalized plasma and cyclotron frequencies must not be
confused with the bare ones defined by the sum rules (\ref{f-sumr})
and (\ref{suhang}), respectively.

Expanding Eqs.\ (\ref{ocond}) and (\ref{hcond}) to leading order in
$1/T$ as explained in Ref.\ \cite{Lange:1998a} shows that both
conductivities are suppressed by a factor $\delta$ close to the Mott
transition. Approximate expressions for the dissipative parts of the
conductivities, which capture the doping and temperature dependences
in the region $T,\omega>T^*$, $\omega\ll2D$, are given by
$\Re\sigma_{xx}(\omega)\sim e^2\delta\,\frac{1-\exp(-|\omega|/T)}
{|\omega|}$ and $\Im\sigma_{xy}(\omega)\sim e^3H\delta\,
\mbox{sgn}(\omega)[1-\exp(-|\omega|/T)]/D$. The last relation only
holds for a generic band that does not have the bipartite-lattice
property discussed in Ref.\ \cite{Lange:1998a}.

We now discuss the qualitative forms of the functions governing the
sum rules (\ref{suhconre}), (\ref{suhconim}), (\ref{acHallre}),
(\ref{acHallim}), and (\ref{suhang}) more specifically. In all
plots, we have chosen $\delta=0.1$ and $U=4$. 

\begin{figure}[ht]
	\epsfig{file=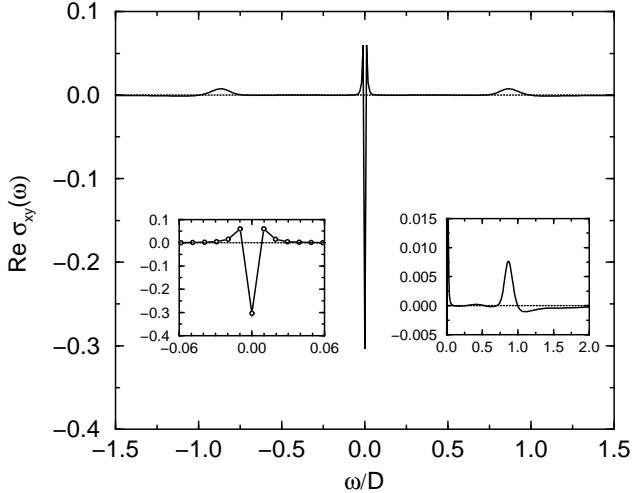,%
	width=2.6in,angle=-90,
	bbllx=87pt,bblly=6pt,bburx=588pt,bbury=638pt}
	\vspace{6pt}
	\caption{Real part of the Hall conductivity for $T=0.015D$. 
	The insets magnify the structures at $\omega=0$ (left 
	one) and $\omega=D$ (right one).} 
\label{fig:hconre1}
\end{figure}

{\it Real part of the Hall conductivity.}--Its high-frequency behavior
is given by $\sigma_{xy}(\omega)\simeq-e^2\chi^0\omega_H/\omega^2$ and
therefore has the opposite sign as $R_H^*$ in Eq.\ (\ref{Ihall}). On
the other hand, its dc value has the same sign as $R_H$. Close to half
filling, and for intermediate temperatures and up, $R_H$ and $R_H^*$
have the same sign \cite{Majumdar:1995}. Since in this parameter
regime, the only energy scale is $D$, $\Re\sigma_{xy}(\omega)$ changes
its sign once at a scale of order $D$ to satisfy sum rule
(\ref{suhconre}). For $T<T_{\mbox{\small coh}}$, $R_H^*$ remains
hole-like while $R_H$ becomes electron-like \cite{Majumdar:1995}. 
Then, the sum rule (\ref{suhconre}) requires at least one further sign
change at a scale $\omega\sim T_{\mbox{\small coh}}$. This prediction
is corroborated by our numerical investigation, Fig.\
\ref{fig:hconre1} displays the frequency-dependent Hall conductivity
for $T=0.015D$.

\begin{figure}[ht]
	\epsfig{file=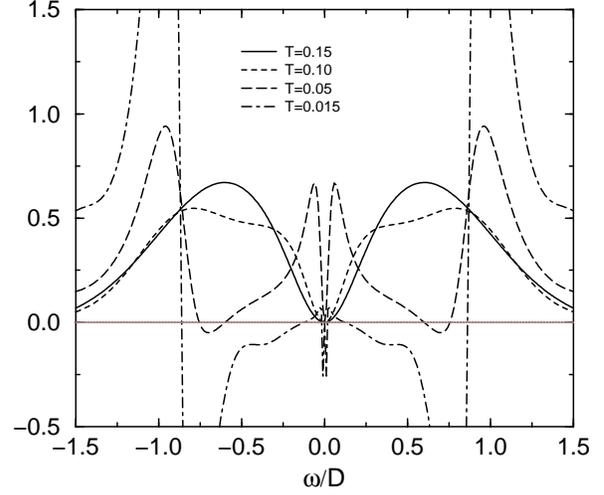,%
	width=2.6in,
	angle=-90,bbllx=87pt,bblly=56pt,bburx=589pt,bbury=638pt}
	\vspace{6pt}
	\caption{$\chi_{xy}''(\omega)$ normalized to 1.}
\label{fig:s_H}
\end{figure}

{\it Imaginary part of the Hall conductivity.}--Fig.\ \ref{fig:s_H}
displays the integrand of the sum rule (\ref{suhconim}), which is
proportional to the spectral function (\ref{chixy}). We have
normalized this function to 1 to facilitate the comparison between
curves belonging to different temperatures. For $T>T^*$, this function
hardly depends on temperature. Its ``M-shaped'' form is consistent
with Eq.\ (\ref{s_H(0)}) and the fact that $D$ is the only energy
scale. As the temperature is decreased to below $T^*$, the spectral
weight is redistributed to comply with the emergence of two energy
scales $T_{\mbox{\small coh}}$ and $\omega_1$, the Drude form in the
Fermi-liquid regime, and the fact that the overall weight is positive.

{\it Hall angle.}--The real part of the Hall angle defined before Eq.\
(\ref{suhang}) closely resembles that of the previously considered
function, except that it is not subject to a condition like Eq.\
(\ref{s_H(0)}). 

\begin{figure}[ht]
	\epsfig{file=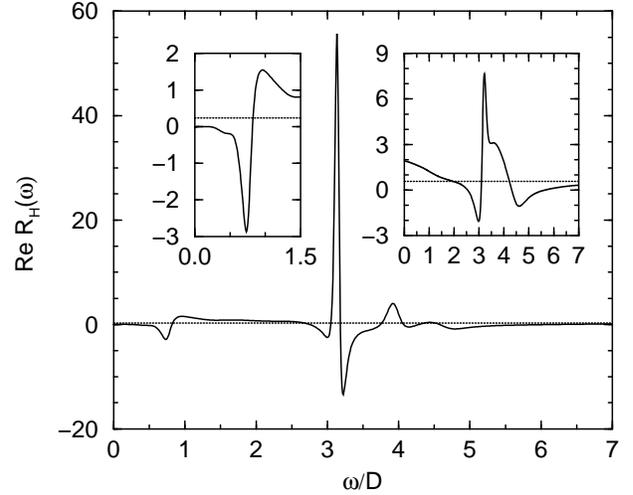,%
	width=2.6in,
	angle=-90,bbllx=88pt,bblly=29pt,bburx=589pt,bbury=638pt}
	\vspace{6pt}
	\caption{$\Re R_H(\omega)$ (solid lines) and $R_H^*$ (dotted
	lines) for $T=0.015D$ (main panel) and $T=0.15D$ (right
	inset). The left inset magnifies the low-frequency
	part of the $T=0.015D$ curve.} 	
\label{fig:RHre}
\end{figure}

{\it Hall constant.}--Close to half filling and for $T>T^*$, $R_H$ is
greater than $R_H^*$ \cite{Majumdar:1995}. Then, $\Re R_H(\omega)$
satisfies the sum rule (\ref{acHallre}) as follows: Starting from its
dc value, $\Re R_H(\omega)$ first decreases monotonically as a
function of frequency, drops to below its infinite-frequency level at
a frequency of order $D$, and finally rises to approach $R_H^*$ from
below. In the opposite limit of very low temperatures, we show a curve
for $T=0.015D$ in the main panel of Fig.\ \ref{fig:RHre}, along with a
better resolution of its low-frequency part in the left inset. $R_H^*$
(dotted line) is seen to be positive, while $R_H<0$ (not discernible),
in agreement with Ref.\ \cite{Majumdar:1995}. In addition,
$R_H(\omega)$ hardly depends on frequency in the Fermi-liquid regime,
as expected from the Drude parametrizations of $\sigma_{xx}(\omega)$
and $\sigma_{xy}(\omega)$. To counterbalance the drop of the dc value
to below $R_H^*$, a peak-like structure has piled up to above the
$R_H^*$ level at the other energy scale, $\omega\sim D$ . The
structure about $\omega\sim 3D$ arises from the upper Hubbard band. At
very high frequencies, $\Re R_H(\omega)$ approaches its asymptotic
value according to a $1/\omega^2$ law. In the right inset of Fig.\
\ref{fig:RHre}, we display a curve at the cross-over temperature
$T=0.15D$. Like in the high-temperature regime, $R_H>R_H^*>0$. But the
sign change of $\Re R_H(\omega)-R_H^*$ is already shifted to higher
frequencies, signalling the emergence of the peak at $\omega\sim D$ as
the temperature is lowered.

\begin{figure}[ht]
	\epsfig{file=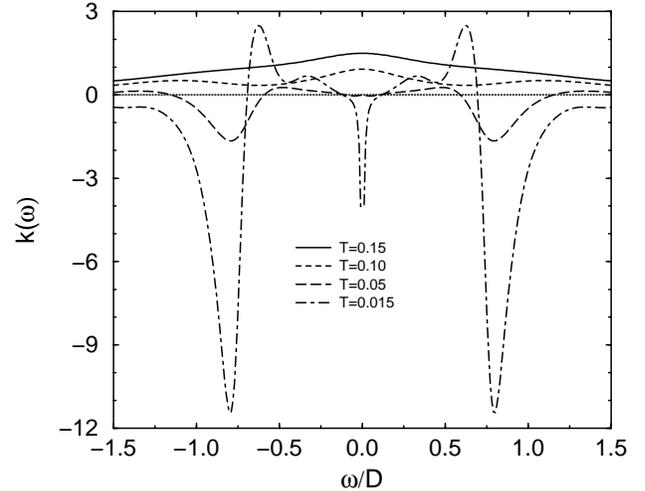,%
	width=2.6in,
	angle=-90,bbllx=87pt,bblly=29pt,bburx=589pt,bbury=638pt}
	\vspace{6pt}
	\caption{$k(\omega)$ defined in Eq.\ (\ref{acRH}).}
\label{fig:kofw}
\end{figure}

Finally, Fig.\ \ref{fig:kofw} displays the function $k(\omega)$, which
is proportional to the integrand of the sum rule (\ref{acHallim}) and
which has not been normalized to 1. As the temperature is decreased
from well above (not shown in Fig.\ \ref{fig:kofw}) to well below
$T^*$, a single peak of width $D$ decomposes into a narrow one at
$\omega=0$ and of width smaller than $T_{\mbox{\small coh}}$, and a
feature around $\omega\sim\omega_1$ which involves a sign change. In
the normal state of cuprates such as La$_{2-x}$Sr$_x$CuO$_4$,
$k(\omega)$ is resonance-like and has a width given by the anomalous
relaxation rate $1/\tau_H$ which exhibits a $T^2$ law
\cite{Lange:1997a}. Instead, we find a width of order $D$ for
temperatures above $T^*$ where $R_H>0$. Similarly, the dynamical
mean-field theory predicts that Kohler's rule is replaced by
$\Delta\rho/\rho\sim(\omega_c/D)^2$ in the high-temperature regime
\cite{Lange:1998a}, whereas experiments on cuprates are consistent
with $\Delta\rho/\rho\sim(\omega_c\tau_H)^2$ as suggested by Terasaki
{\it et al.} \cite{Terasaki:1995}. Here, $\Delta\rho$ is the
magnetoresistance.

In summary, we have derived sum rules for the real and imaginary parts
of the Hall conductivity and Hall constant. We have applied them, along
with another one for the Hall angle, to the doped Mott insulator.

This work was supported by NSF DMR 95-29138. E.L. is funded by the
Deutsche Forschungsgemeinschaft.

%%%%%%%%%%%%%%%%%%%%%%%%%%%%%%%%%%%%%%%%%%%%%%%%%%%%%%%%%%%%%%%%%%%%%%%%%%%%%
%%%%%%   
%%%%%%           FIGURE CAPTIONS
%%%%%%   
%%%%%%%%%%%%%%%%%%%%%%%%%%%%%%%%%%%%%%%%%%%%%%%%%%%%%%%%%%%%%%%%%%%%%%%%%%%%%

%\newpage

%----------------------------------------------------------------------------
%----------------------------------------------------------------------------
%----------------------------------------------------------------------------
%----------------------------------------------------------------------------
%----------------------------------------------------------------------------

\end{document}